\documentclass[12pt, graphicx]{article}
\usepackage{hyperref}

\def\be{\begin{equation}}
\def\ee{\end{equation}}
\def\bea{\begin{eqnarray}}
\def\eea{\end{eqnarray}}

\def\dt#1{\on{\hbox{\bf .}}{#1}}                
\def\Dot#1{\dt{#1}}
\def\IR{\relax{\rm I\kern-.18em R}}
\def\binomial#1#2{\left(\,{\buildrel 
{\raise4pt\hbox{$\displaystyle{#1}$}}\over
{\raise-6pt\hbox{$\displaystyle{#2}$}}}\,\right)}

\def\[{\lfloor{\hskip 0.35pt}\!\!\!\lceil}
\def\]{\rfloor{\hskip 0.35pt}\!\!\!\rceil}

\newcommand{\AmS}{{\protect\the\textfont2
  A\kern-.1667em\lower.5ex\hbox{M}\kern-.125emS}}

\catcode`@=11
\def\un#1{\relax\ifmmode\@@underline#1\else
        $\@@underline{\hbox{#1}}$\relax\fi}
\catcode`@=12

\def\ad{{\kern0.5pt
                   \alpha \kern-5.05pt
\raise5.8pt\hbox{$\textstyle.$}\kern
0.5pt}}

\def\Dot#1{{\kern0.5pt
     {#1} \kern-5.05pt \raise5.8pt\hbox{$\textstyle.$}\kern
0.5pt}}

\def\a{\alpha}
\def\b{\beta}

\def\d{\delta}

\def\g{\gamma}

\def\k{\kappa}

\def\q{\theta}

\def\s{\sigma}

\def\z{\zeta}

\def\O{\Omega}
\def\P{\Pi}

\def\S{\Sigma}
\def\U{\Upsilon}

\def\cg{{\cal G}}

\def\cs{{\cal S}}

\def\cu{{\cal U}}

\def\bo{{\raise.15ex\hbox{\large$\Box$}}}               
\def\pa{\partial}                                       
\def\TH{{\raise.2ex\hbox{$\displaystyle \bigodot$}\mskip-4.7mu \llap H
\;}}
\def\face{{\raise.2ex\hbox{$\displaystyle \bigodot$}\mskip-2.2mu \llap
{$\ddot
        \smile$}}}                                      

\def\Bar#1{\overline{#1}}                       
\def\leftrightarrowfill{$\mathsurround=0pt \mathord\leftarrow \mkern-6mu
        \cleaders\hbox{$\mkern-2mu \mathord- \mkern-2mu$}\hfill
        \mkern-6mu \mathord\rightarrow$}
\def\dvec#1{\vbox{\ialign{##\crcr
        \leftrightarrowfill\crcr\noalign{\kern-1pt\nointerlineskip}
        $\hfil\displaystyle{#1}\hfil$\crcr}}}           
\def\dt#1{{\buildrel {\hbox{\LARGE .}} \over {#1}}}     

\def\frac#1#2{{\textstyle{#1\over\vphantom2\smash{\raise.20ex
        \hbox{$\scriptstyle{#2}$}}}}}                   
\def\sfrac#1#2{{\vphantom1\smash{\lower.5ex\hbox{\small$#1$}}\over
        \vphantom1\smash{\raise.4ex\hbox{\small$#2$}}}} 
\def\bfrac#1#2{{\vphantom1\smash{\lower.5ex\hbox{$#1$}}\over
        \vphantom1\smash{\raise.3ex\hbox{$#2$}}}}       
\def\afrac#1#2{{\vphantom1\smash{\lower.5ex\hbox{$#1$}}\over#2}}    

\newskip\humongous \humongous=0pt plus 1000pt minus 1000pt

\newif\ifdtup

  \def\pp{{\mathchoice
          {
              \kern 1pt%
              \raise 1pt
              \vbox{\hrule width5pt height0.4pt depth0pt
                    \kern -2pt
                    \hbox{\kern 2.3pt
                          \vrule width0.4pt height6pt depth0pt
                          }
                    \kern -2pt
                    \hrule width5pt height0.4pt depth0pt}%
                    \kern 1pt
           }
            {
              \kern 1pt%
              \raise 1pt
              \vbox{\hrule width4.3pt height0.4pt depth0pt
                    \kern -1.8pt
                    \hbox{\kern 1.95pt
                          \vrule width0.4pt height5.4pt depth0pt
                          }
                    \kern -1.8pt
                    \hrule width4.3pt height0.4pt depth0pt}%
                    \kern 1pt
            }
            {
              \kern 0.5pt%
              \raise 1pt
              \vbox{\hrule width4.0pt height0.3pt depth0pt
                    \kern -1.9pt  
                    \hbox{\kern 1.85pt
                          \vrule width0.3pt height5.7pt depth0pt
                          }
                    \kern -1.9pt
                    \hrule width4.0pt height0.3pt depth0pt}%
                    \kern 0.5pt
            }
            {
              \kern 0.5pt%
              \raise 1pt
              \vbox{\hrule width3.6pt height0.3pt depth0pt
                    \kern -1.5pt
                    \hbox{\kern 1.65pt
                          \vrule width0.3pt height4.5pt depth0pt
                          }
                    \kern -1.5pt
                    \hrule width3.6pt height0.3pt depth0pt}%
                    \kern 0.5pt
            }
        }}

  \def\mm{{\mathchoice
                       {
                             \kern 1pt
               \raise 1pt    \vbox{\hrule width5pt height0.4pt depth0pt
                                  \kern 2pt
                                  \hrule width5pt height0.4pt depth0pt}
                             \kern 1pt}
                       {
                            \kern 1pt
               \raise 1pt \vbox{\hrule width4.3pt height0.4pt depth0pt
                                  \kern 1.8pt
                                  \hrule width4.3pt height0.4pt depth0pt}
                             \kern 1pt}
                       {
                            \kern 0.5pt
               \raise 1pt
                            \vbox{\hrule width4.0pt height0.3pt depth0pt
                                  \kern 1.9pt
                                  \hrule width4.0pt height0.3pt depth0pt}
                            \kern 1pt}
                       {
                           \kern 0.5pt
             \raise 1pt  \vbox{\hrule width3.6pt height0.3pt depth0pt
                                  \kern 1.5pt
                                  \hrule width3.6pt height0.3pt depth0pt}
                           \kern 0.5pt}
                       }}

\def\pd{{\kern0.5pt
                   + \kern-5.05pt \raise5.8pt\hbox{$\textstyle.$}\kern
0.5pt}}

\def\pmd{{\kern0.5pt
                  \pm \kern-5.05pt \raise6.3pt\hbox{$\textstyle.$}\kern1.5pt}}

\def\md{{\mathchoice
   {
      {{\kern 1pt - \kern-6.2pt \raise5pt\hbox{$\textstyle.$}\kern 1pt}}}
    {
      {{\kern 1pt - \kern-6.2pt \raise5pt\hbox{$\textstyle.$}\kern 1pt}}}
    {
      {\kern0.5pt - \kern-5.05pt \raise3.4pt\hbox{$\textstyle.$}\kern0.5pt}}
    {
      {\kern0.5pt - \kern-5.05pt \raise3.4pt\hbox{$\textstyle.$}\kern0.5pt}}}}

\def\ad{{\dot{\alpha}}}
\def\bd{{\dot{\beta}}}

\def\pp{{\mathchoice
          {
              \kern 1pt%
              \raise 1pt
              \vbox{\hrule width5pt height0.4pt depth0pt
                    \kern -2pt
                    \hbox{\kern 2.3pt
                          \vrule width0.4pt height6pt depth0pt
                          }
                    \kern -2pt
                    \hrule width5pt height0.4pt depth0pt}%
                    \kern 1pt
           }
            {
              \kern 1pt%
              \raise 1pt
              \vbox{\hrule width4.3pt height0.4pt depth0pt
                    \kern -1.8pt
                    \hbox{\kern 1.95pt
                          \vrule width0.4pt height5.4pt depth0pt
                          }
                    \kern -1.8pt
                    \hrule width4.3pt height0.4pt depth0pt}%
                    \kern 1pt
            }
            {
              \kern 0.5pt%
              \raise 1pt
              \vbox{\hrule width4.0pt height0.3pt depth0pt
                    \kern -1.9pt  
                    \hbox{\kern 1.85pt
                          \vrule width0.3pt height5.7pt depth0pt
                          }
                    \kern -1.9pt
                    \hrule width4.0pt height0.3pt depth0pt}%
                    \kern 0.5pt
            }
            {
              \kern 0.5pt%
              \raise 1pt
              \vbox{\hrule width3.6pt height0.3pt depth0pt
                    \kern -1.5pt
                    \hbox{\kern 1.65pt
                          \vrule width0.3pt height4.5pt depth0pt
                          }
                    \kern -1.5pt
                    \hrule width3.6pt height0.3pt depth0pt}%
                    \kern 0.5pt
            }
        }}

  \def\mm{{\mathchoice
                       {
                             \kern 1pt
               \raise 1pt    \vbox{\hrule width5pt height0.4pt depth0pt
                                  \kern 2pt
                                  \hrule width5pt height0.4pt depth0pt}
                             \kern 1pt}
                       {
                            \kern 1pt
               \raise 1pt \vbox{\hrule width4.3pt height0.4pt depth0pt
                                  \kern 1.8pt
                                  \hrule width4.3pt height0.4pt depth0pt}
                             \kern 1pt}
                       {
                            \kern 0.5pt
               \raise 1pt
                            \vbox{\hrule width4.0pt height0.3pt depth0pt
                                  \kern 1.9pt
                                  \hrule width4.0pt height0.3pt depth0pt}
                            \kern 1pt}
                       {
                           \kern 0.5pt
             \raise 1pt  \vbox{\hrule width3.6pt height0.3pt depth0pt
                                  \kern 1.5pt
                                  \hrule width3.6pt height0.3pt depth0pt}
                           \kern 0.5pt}
                       }}

\def\pd{{\kern0.5pt
                   + \kern-5.05pt \raise5.8pt\hbox{$\textstyle.$}\kern
0.5pt}}

\def\pmd{{\kern0.5pt
                  \pm \kern-5.05pt \raise6.3pt\hbox{$\textstyle.$}\kern1.5pt}}

\def\md{{\mathchoice
   {
      {{\kern 1pt - \kern-6.2pt \raise5pt\hbox{$\textstyle.$}\kern 1pt}}}
    {
      {{\kern 1pt - \kern-6.2pt \raise5pt\hbox{$\textstyle.$}\kern 1pt}}}
    {
      {\kern0.5pt - \kern-5.05pt \raise3.4pt\hbox{$\textstyle.$}\kern0.5pt}}
    {
      {\kern0.5pt - \kern-5.05pt \raise3.4pt\hbox{$\textstyle.$}\kern0.5pt}}}}

\def\dslash{\not{\hbox{\kern-2pt $\partial$}}}
\def\Dslash{\not{\hbox{\kern-4pt $D$}}}
\def\pslash{\not{\hbox{\kern-2.3pt $p$}}}
 \newtoks\slashfraction
 \slashfraction={.13}
 \def\slash#1{\setbox0\hbox{$ #1 $}
 \setbox0\hbox to \the\slashfraction\wd0{\hss \box0}/\box0 }
 
\font\ro=cmsy10                          
\def\kcr{{\hbox{\ro \char'170}}}                
\def\ktl{{\hbox{\ro \char'170}}}        
\def\ktr{{\hbox{\ro \char'170}}}        
\def\kbl{{\hbox{\ro \char'170}}}        
\def\kbr{{\hbox{\ro \char'170}}}        

\def\plpl{\raise-2pt\hbox{$\raise3pt\hbox{$_+$}\hskip-6.67pt\raise0.0pt
\hbox{$^+$}\hskip 0.01pt$}}
\def\mimi{\raise-2pt\hbox{$\raise3pt\hbox{$_-$}\hskip-6.67pt\raise0.0pt
\hbox{$^-$}\hskip 0.01pt$}} 

\def\bo{{\raise.15ex\hbox{\large$\Box$}}}               
\def\pa{\partial}                                       
\def\TH{{\raise.2ex\hbox{$\displaystyle \bigodot$}\mskip-4.7mu \llap H \;}}
\def\face{{\raise.2ex\hbox{$\displaystyle \bigodot$}\mskip-2.2mu \llap {$\ddot
        \smile$}}}                                      

\def\Bar#1{\overline{#1}}                       
\def\leftrightarrowfill{$\mathsurround=0pt \mathord\leftarrow \mkern-6mu
        \cleaders\hbox{$\mkern-2mu \mathord- \mkern-2mu$}\hfill
        \mkern-6mu \mathord\rightarrow$}
\def\dvec#1{\vbox{\ialign{##\crcr
        \leftrightarrowfill\crcr\noalign{\kern-1pt\nointerlineskip}
        $\hfil\displaystyle{#1}\hfil$\crcr}}}           
\def\dt#1{{\buildrel {\hbox{\LARGE .}} \over {#1}}}     

\def\frac#1#2{{\textstyle{#1\over\vphantom2\smash{\raise.20ex
        \hbox{$\scriptstyle{#2}$}}}}}                   
\def\sfrac#1#2{{\vphantom1\smash{\lower.5ex\hbox{\small$#1$}}\over
        \vphantom1\smash{\raise.4ex\hbox{\small$#2$}}}} 
\def\bfrac#1#2{{\vphantom1\smash{\lower.5ex\hbox{$#1$}}\over
        \vphantom1\smash{\raise.3ex\hbox{$#2$}}}}       
\def\afrac#1#2{{\vphantom1\smash{\lower.5ex\hbox{$#1$}}\over#2}}    

\renewcommand{\Box}{\,\raisebox{-.45pt}{\drawsquare{6}{0.6}}\,}

\newcommand{\drawsquare}[2]{\hbox{%

\rule{#2pt}{#1pt}\hskip-#2pt

\rule{#1pt}{#2pt}\hskip-#1pt

\rule[#1pt]{#1pt}{#2pt}}\rule[#1pt]{#2pt}{#2pt}\hskip-#2pt

\rule{#2pt}{#1pt}}

\parskip=\medskipamount                 
\thicklines                         

\def\oldheadpic{                                
        \setlength{\unitlength}{.4mm}
        \thinlines
        \par
        \begin{picture}(349,16)
        \put(325,16){\line(1,0){4}}
        \put(330,16){\line(1,0){4}}
        \put(340,16){\line(1,0){4}}
        \put(335,0){\line(1,0){4}}
        \put(340,0){\line(1,0){4}}
        \put(345,0){\line(1,0){4}}
        \put(329,0){\line(0,1){16}}
        \put(330,0){\line(0,1){16}}
        \put(339,0){\line(0,1){16}}
        \put(340,0){\line(0,1){16}}
        \put(344,0){\line(0,1){16}}
        \put(345,0){\line(0,1){16}}
        \put(329,16){\oval(8,32)[bl]}
        \put(330,16){\oval(8,32)[br]}
        \put(339,0){\oval(8,32)[tl]}
        \put(345,0){\oval(8,32)[tr]}
        \end{picture}
        \par
        \thicklines
        \vskip.2in}
\def\oldtitle#1#2#3#4{\oldheadpic\begin{center}\vglue.5in{\large\bf #1}\\[.6in]
        {#2}\\[.1in] {\it Department of Physics and Astronomy}\\
        {\it University of Maryland, College Park, MD 20742}\\[.6in]
        Physics Publication \#{#3}\\ {#4}\\[1.5in] {\bf ABSTRACT}\\[.1in]
        \end{center} \begin{quotation}}                 
\def\oldTitle#1#2#3#4#5#6#7{\oldheadpic\begin{center} \vglue .4in
        {\large\bf #1}\\[.4in]
        {#2}\\[.1in] {\it Department of Physics and Astronomy}\\
        {\it University of Maryland, College Park, MD 20742}\\[.1in]
        {#3}\\[.1in] {\it {#4}}\\ {\it {#5}}\\[.4in]
        Physics Publication \#{#6}\\ {#7}\\[.5in] {\bf ABSTRACT}\\[.1in]
        \end{center} \begin{quotation}}                 
\def\border{                                            
        \setlength{\unitlength}{1mm}
        \newcount\xco
        \newcount\yco
        \xco=-21
        \yco=12
        \begin{picture}(140,0)
        \put(\xco,\yco){$\ktl$}
        \advance\yco by-1
        {\loop
        \put(\xco,\yco){$\kcr$}
        \advance\yco by-2
        \ifnum\yco>-240
        \repeat
        \put(\xco,\yco){$\kbl$}}
        \xco=158
        \yco=12
        \put(\xco,\yco){$\ktr$}
        \advance\yco by-1
        {\loop
        \put(\xco,\yco){$\kcr$}
        \advance\yco by-2
        \ifnum\yco>-240
        \repeat
        \put(\xco,\yco){$\kbr$}}
        \put(-20,13){\tiny  University of Maryland ** Center for String and 
         Particle  Theory ** Physics Department * University of Maryland ** Center  
        for String and Particle  Theory}
        \put(-20,-241.5){\tiny  University of Maryland ** Center for String and 
         Particle  Theory ** Physics Department * University of Maryland ** Center  
        for String and Particle  Theory}        \end{picture}
        \par\vskip-8mm}
\def\bordero{                                           
        \setlength{\unitlength}{1mm}
        \newcount\xco
        \newcount\yco
        \xco=-31
        \yco=12
        \begin{picture}(140,0)
        \put(\xco,\yco){$\ktl$}
        \advance\yco by-1
        {\loop
        \put(\xco,\yco){$\kclr}
        \advance\yco by-2
        \ifnum\yco>-240
        \repeat
        \put(\xco,\yco){$\kbl$}}
        \xco=151
        \yco=12
        \put(\xco,\yco){$\ktr$}
        \advance\yco by-1
        {\loop
        \put(\xco,\yco){$\kcr$}
        \advance\yco by-2
        \ifnum\yco>-240
        \repeat
        \put(\xco,\yco){$\kbr$}}
        \put(-20,12){\ooo bacdefghidfghghdhededbihdgdfdfhhdheidhdhebaaahjhhdahba

hgdedge
   hgfdiehhgdigicba}
        \put(-20,-241.5){\ooo ababaighefdbfghgeahgdfgafagihdidihiidhiagfedhadbfd

ecdcdfa
   gdcbhaddhbgfchbgfdacfediacbabab}
        \end{picture}
        \par\vskip-8mm}
\def\headpic{                                           
        \indent
        \setlength{\unitlength}{.4mm}
        \thinlines
        \par
        \begin{picture}(29,16)
        \put(165,16){\line(1,0){4}}
        \put(170,16){\line(1,0){4}}
        \put(180,16){\line(1,0){4}}
        \put(175,0){\line(1,0){4}}
        \put(180,0){\line(1,0){4}}
        \put(185,0){\line(1,0){4}}
        \put(169,0){\line(0,1){16}}
        \put(170,0){\line(0,1){16}}
        \put(179,0){\line(0,1){16}}
        \put(180,0){\line(0,1){16}}
        \put(184,0){\line(0,1){16}}
        \put(185,0){\line(0,1){16}}
        \put(169,16){\oval(8,32)[bl]}
        \put(170,16){\oval(8,32)[br]}
        \put(179,0){\oval(8,32)[tl]}
        \put(185,0){\oval(8,32)[tr]}
        \end{picture}
        \par\vskip-6.5mm
        \thicklines}
\def\title#1#2#3#4{\border\headpic {\hbox to\hsize{#4 \hfill UMDEPP #3}}\par
        \begin{center} \vglue .5in {\large\bf #1}\\[.6in]
        {#2}\\[.1in] {\it Department of Physics and Astronomy}\\
        {\it University of Maryland, College Park, MD 20742}\\[1.5in]
        {\bf ABSTRACT}\\[.1in] \end{center} \begin{quotation}}  
\def\Title#1#2#3#4#5#6#7{\border\headpic
        {\hbox to\hsize{#7 \hfill UMDEPP #6}}\par
        \begin{center} \vglue .4in {\large\bf #1}\\[.4in]
        {#2}\\[.1in] {\it Department of Physics and Astronomy}\\
        {\it University of Maryland, College Park, MD 20742}\\[.1in]
        {#3}\\[.1in] {\it {#4}}\\ {\it {#5}}\\[.5in] {\bf ABSTRACT}\\[.1in]
        \end{center} \begin{quotation}}                 
\def\endtitle{\end{quotation}\newpage}                  

\def\qd{{\kern0.5pt
                   q \kern-5.05pt \raise5.8pt\hbox{$\textstyle.$}\kern
0.5pt}}

\newcommand{\ve}{\varepsilon}

\begin{document}

\def\gfrac#1#2{\frac {\scriptstyle{#1}}
         {\mbox{\raisebox{-.6ex}{$\scriptstyle{#2}$}}}}
\def\gg{{\hbox{\sc g}}}
\border\headpic {\hbox to\hsize{June 2003 \hfill
{UMDEPP 03-053}}}
\par
\setlength{\oddsidemargin}{0.3in}
\setlength{\evensidemargin}{-0.3in}
\begin{center}
\vglue .10in
{\large\bf The Off-Shell (3/2, 2) Supermultiplets
Revisited\footnote
{Supported in part  by National Science Foundation Grant
PHY-0099544, the Australian 
\newline ${~~~~\,}$ Research Council, the Australian
Academy of Science as well as by UWA research grants.}\  }
\\[.5in]
S. James Gates, Jr.\footnote{gatess@wam.umd.edu}${}^\star$,  Sergei M.
Kuzenko\footnote{kuzenko@cyllene.uwa.edu.au}${}^\ast$ and J.
Phillips\footnote{ferrigno@physics.umd.edu}${}^\star$
\\[0.06in]
{\it ${}^\star$Center for String and Particle Theory\\
Department of Physics, University of Maryland\\
College Park, MD 20742-4111 USA}\\[0.06in]
{\it ${}^\ast$ School of Physics, The University of Western Australia\\
Crawley, W.A. 6009, Australia}\\[1.5in]

{\bf ABSTRACT}\\[.01in]
\end{center}
\begin{quotation}
{Using superspace projection operators we
provide a classification of  $(3/2,2)$ off-shell
supermultiplets which are realized in terms of a real
axial vector superfield, with or without compensating
superfields. Any linearized supergravity action is shown
to be a  superposition of those corresponding to
(i) old minimal  supergravity,  (ii) new minimal supergravity
and (iii) the novel  $(3/2,2)$ off-shell supermultiplet with
$12+12$   degrees of freedom obtained in hep-th/0201096.}

${~~~}$ \newline

PACS: 04.65.+e, 11.15.-q, 11.25.-w, 12.60.J
\endtitle
\section{Introduction and Outlook }
~~~~Different off-shell realizations of four-dimensional  ${\cal N}=1$ 
supergravity and their couplings to supersymmetric matter were thoroughly 
studied in the late 1970s -- early 1980s,  and by now these issues have
become a subject of two comprehensive textbooks \cite{Gates1,Buch1}
(see also \cite{Wess} for an introduction to old minimal supergravity). 
Therefore one could hardly expect to say anything new about off-shell 
${\cal N}=1$ supergravity.  A surprise came a year ago.  While investigating 
superfield models for the massive superspin-$\frac 32$ multiplet,\footnote{See also \cite{Buch3}  for the study of off-shell realizations of the massive 
gravitino multiplet.}  Buchbinder et al. \cite{Buch2} found a new version 
of linearized supergravity when setting the mass parameter  to zero.  
To the best of our knowledge, this  model  had been overlooked in all 
previous investigations.  Its existence indicates that along with the old 
minimal, new minimal and non-minimal supergravity formulations,
there may exist a new realization, with possibly interesting properties.
Its existence also naturally calls upon working out a classification of 
all free  $(3/2, 2)$ supermultiplets.  The present note is aimed at deriving 
such a classification.  An unexpected outcome of our analysis is that the 
novel formulation of \cite{Buch2} taken together with the old and new 
minimal formulations are the main building blocks for generating all 
possible linearized supergravity actions.

Our approach  in this paper is  very similar,  in spirit, to the work  \cite{GS}
which provided the classification of minimal free $(1, 3/2)$ supermultiplets 
(massless gravitino multiplet) in terms of an unconstrained spinor superfield 
$\Psi_\a$ and its conjugate ${\bar \Psi}_\ad$.  This was accomplished via 
the use of ${\cal N}=1$ superfield projector operators  \cite{Sok,SG} as 
formulated in \cite{SG}.  In the case of massless $(3/2, 2)$ supermultiplet,
we are going to  consider a general  linearized {\it local}  action for a real 
axial vector superfield $H_{\a \ad}$ and then rewrite it in terms of relevant 
superprojectors. Gauge invariant models emerge if one requires that some 
superprojectors are not present in the action.  As is  shown below, there  
exist gauge invariant actions with two, three and four superprojectors present. 
Not all of such actions, however, describe a pure linearized supergravity
multiplet. Some of them may represent a particular coupling of linearized 
supergravity to supersymmetric matter, and  these should be discarded.
The most tedious part of the classification problem is to select those models 
which indeed describe a single massless $(3/2, 2)$ supermultiplet.

One of the main motivations for the present work was the desire to 
achieve a better understanding of supersymmetric higher spin
multiplets, both in the  massless and massive cases.  Superstring 
theory predicts the existence of an infinite tower of higher spins 
supermultiplets, and therefore it  seems important to have their 
manifestly supersymmetric field theoretic description(s).  On the 
other hand, massless higher spin supermultiplets are of some 
importance in the framework of the AdS/CFT correspondence (see, 
e.g. \cite{Sundborg,Witten,Vasiliev,Mikhailov,Tseytlin,GKP}
and references therein). As concerns the massive case, off-shell 
higher spin supermultiplets have never been constructed.  In the 
massless case, for each superspin $s> 3/2$ there exist two dually 
equivalent off-shell realizations in 4D ${\cal N}=1$ flat superspace 
\cite{KS1} (see \cite{Buch1} for a review) and anti-de Sitter superspace 
\cite{KS2}. We also know the structure of 4D ${\cal N}=2$  off-shell
higher spin supermultiplets \cite{GKS1} as well as a generating
superfield action for arbitrary superspin massless multiplets in 
4D ${\cal N}=1$ anti-de Sitter superspace \cite{GKS2}.  One thing
still missing is a classification of massless higher spin supermultiplets 
(say, those involving, for a half-integer superspin $s =p +1/2$,
 a real tensor superfield $H_{(\a_1\cdots\a_p) (\ad_1 \cdots \ad_p)}$ 
along with  some compensators).

In the case of half-integer superspin, for $s=3/2$ one of the families constructed in  \cite{KS1} reduces to linearized old minimal supergravity, while  the other -- to linearized $n=-1$ non-minimal supergravity.  Is there a series of massless higher spin supermultiplets\footnote{Recently, Engquist et al. \cite{ESS} have formulated, following generalizations of the approach 
\newline ${~~~~\,}$ pioneered in \cite{FV},  nonlinear equations for interacting massless higher spin ${\cal N}=1$ super-
\newline${~~~~\,}$ fields in four space-time dimensions.  No analysis was given, however,  as to the relation- 
\newline ${~~~~\,}$ ship between the spectrum of their model  at 
the linearized level and the free massless 
\newline ${~~~~\,}$ supermultiplets in 4D ${\cal N}=1$ anti-de Sitter superspace introduced  in \cite{KS2, GKS1,GKS2} .}
that terminates at linearized new minimal supergravity (or the novel 
formulation given in \cite{Buch2}) for  $s=3/2$? Presently, we do not know 
an answer to this question. Since superstring compactifications 
are often believed to favor the new minimal formulation of  supergravity, the question does not seem to be of purely academic interest.  We plan to address this and related  questions in a future publication. There are some grounds to 
believe that the analysis undertaken in the present note, can be 
generalized to provide a classification of higher spin massless 
supermultiplets.

This note is  organized as follows.  Section 2 is devoted to describing
the superprojector setup which is crucial for our subsequent analysis.  
In section 3 we derive all minimal gauge invariant actions which involve 
two superprojectors. In section 4 the models with three projectors are 
analyzed.  We demonstrate that all such models describe linearized
supergravity coupled to a scalar multiplet.  Finally, in section 4 we 
consider gauge invariant actions with four projectors and show that they 
describe linearized non-minimal supergravity parameterized by a complex 
parameter $n$. The case of real $n$ is  studied in detail.
\section{Setup}
We start with the most general  linearized  action for  $H_{\un a}$  required 
to be local, CPT even and of fourth order in  spinor derivatives:
\bea
\label{rawaction}
{\cal S} ~=~ \int d^8z \Big\{\a_1 H^{\un a}D^\b\Bar D^2D_\b H_{\un a}
&+&\a_2 H^{\un a}\Box H_{\un a}~~~~~~~~~~\cr
+~\a_3 H^{\un a}\pa_{\un a}\pa^{\un b}H_{\un b}
&+&\a_4H^{\un a}[D_\a,\Bar D_{\dot\a}][D_{\b},\Bar D_{\dot\b}]H^{\un b}
\Big\}~,
\eea
with $\a_1,\dots,\a_4$ constant dimensionless parameters.  The gravitational 
superfield can then be represented as a superposition of SUSY irreducible 
components,
\bea
 H_{\un a} =  \Big(\P^L_{0}+\P^L_{1/2} +\P^T_{1}+\P^T_{1/2}+\P^T_{3/2} 
\Big)H_{\un a}~,
\eea
by making use of the relevant superprojectors\footnote{We use a slightly 
simplified notation for the superprojectors as compared with \cite{SG}. The 
\newline
${~~~~\,}$ precise dictionary is the following:
$\Pi^L_0 = \Pi^L_{0,0} + \Pi^L_{2,0}$,  $\Pi^L_{1/2} =\Pi^L_{1,1/2}$,
$\Pi^T_{1/2} = \Pi^T_{1,1/2}$,  \newline
${~~~~\,}$ $\Pi^T_1 = \Pi^T_{0,1} + \Pi^T_{2,1}$
and $\Pi^T_{3/2} =\Pi^T_{1,3/2}$.}
\cite{SG,Gates1}
\bea
\P^L_{0}H_{\un a}&=&-{\frac 1{32}}\Box^{{}_{-2}}\pa_{\un a}\{ D^2,
\Bar D^2 \}\pa_{\un c}H^{\un c} ~,\cr
\P^L_{1/2}H_{\un a}&=&
{\frac 1{16}}\Box^{{}_{-2}}\pa_{\un a}D^\d\Bar D^2D_\d \pa_{\un
c}H^{\un c}~, \cr
\P^T_{1/2}H_{\un a}&=&{\frac 1{3!8}}\Box^{{}_{-2}}\pa_{\dot\a}^{~\b}
[D_\b \Bar D^2D^\d \pa_{(\a}^{~~\dot\b}H_{\d)\dot\b}+D_\a\Bar D^2D^\d
\pa_{(\b}^{~~\dot\b}H_{\d)\dot\b}]~, \cr
\P^T_{1}H_{\un a}&=&{\frac 1{32}}\Box^{{}_{-2}}\pa_{\dot\a}^{~\b}\{
D^2, \Bar D^2\} \pa_{(\a}^{~~\dot\b}H_{\b)\dot\b} ~,\cr
\P^T_{3/2}H_{\un a}&=&
-{\frac 1{3!8}}\Box^{{}_{-2}}\pa_{\dot\a}^{~\b} D^\g\Bar D^2
D_{(\g}\pa_\a^{~\dot\b}H_{\b)\dot\b}~.
\eea
Here the superscripts $L$ and $T$ denote longitudinal and transverse
projectors, while the subscripts $0, 1/2, 1, 3/2$ stand for  superspin.
One can readily express the action in terms of the superprojectors.
It is a  D-algebra exercise to show
\bea
D^\g\Bar D^2D_\g H_{\un a} &=&
-8 \Box (\P^L_{1/2}+\P^T_{1/2}
+\P^T_{3/2})H_{\un a}~, \cr
\pa_{\un a}\pa^{\un b}H_{\un b}&=&
-2\Box ( \P^L_{0} +\P^L_{1/2})H_{\un a}~, \cr
\label{doubletrouble}
[D_\a ,\Bar D_{\dot\a}][D_\b ,\Bar D_{\dot\b}]H^{\un  b}&=&
+ \Box (8\P_{0}^{L} -24\,\P^T_{1/2}) H_{\un a}~,
\eea
and therefore the action (\ref{rawaction}) takes the form
\bea
\label{projectedaction}
{\cal S} ~=~ \int d^8z  H^{\un a}\Box \Big\{ (\a_2-2\a_3+8\a_4) \, \P_{0}^L
+(-8\a_1+\a_2-2\a_3) \, \P_{1/2}^L\cr
+ \a_2 \,\P_{1}^T+(-8\a_1+\a_2-24\a_4) \, \P_{1/2}^T
+(-8\a_1+\a_2)\P_{3/2}^T \Big\}H_{\un a}~.
\eea
It  is the projector $\P_{3/2}^T$ which singles out the superspin-3/2 part 
of $H_{\un a}$, and the corresponding projection $\P^T_{ 3/2} \, H_{\un a}$ 
is invariant under the gauge transformation
\bea
\label{gauge}
\d H_{\un a} = \Bar D_{\dot\a}L_{\a}
-D_\a\Bar L_{\dot\a}~,
\eea
which is typical of  linearized conformal supergravity. This also means that
the coefficient multiplying this projector in the action must never vanish.

However, since $\P^T_{3/2}$ is non-local, any local action involving this projector must contain at least one other projector.  The other projectors do not respect the gauge freedom (\ref{gauge}) with  unconstrained $L_\a$, as is seen  from explicit variations
\bea
\label{chil}
\d \int d^8z \, H^{\un a}\Box\P^L_{0}H_{\un a}
&=&+\frac i4  \int d^8z \, \pa_{\un a}H^{\un a} ~ \Big[\Bar D^2D^\a L_\a
-D^2\Bar D_{\dot\a}\Bar L^{\dot\a}\Big];\\
\label{linl}
\d \int d^8z \,
H^{\un a}\Box\P^L_{1/2}H_{\un a}
&=&-\frac i4   \int d^8z \, \pa_{\un a}H^{\un a}~ \Big[D^\a\Bar D^2L_\a
-\Bar D_{\dot\a}D^2\Bar L^{\dot\a}\Big];\\
\label{lint}
\d \int d^8z \, H^{\un a}\Box\P^T_{1/2}H_{\un a}
&=&\frac 18   \int d^8z \, [D_\a, \Bar D_{\dot\a}]H^{\un a}~
\Big[ D^\a\Bar D^2L_\a+\Bar D_{\dot\a}D^2\Bar L^{\dot\a}\Big]; \\
\label{chit}
\d \int d^8z \, H^{\un a}\Box\P^T_{1}H_{\un a}
&=&-\frac i4  \int d^8z \, \Big[\pa_{\dot\a}^{~\b}H^{\un a}\Bar D^2D_{(\a}L_{\b)}
+\pa_{\a}^{~\dot\b}H^{\un a}D^2\Bar D_{(\dot\a}\Bar L_{\dot\b)}\Big]~.
\qquad {}
\eea
Therefore, given a gauge invariant action of the form (\ref{projectedaction}),
with $\a_2 -8\a_1 \neq 0$,  the gauge parameter $L_\a$ in (\ref{gauge})
should obey some constraints.  The gauge freedom thus becomes model
dependent. This  may be avoided at the cost of introducing special lower spin 
compensator(s)   $\U$ with a nontrivial  transformation law under   (\ref{gauge})
such that imposing the {\it admissible} gauge condition $\U = 0$ is equivalent 
to restoring the original constraints on the gauge parameter.  This is of course the standard compensator philosophy in supersymmetric theories \cite{Gates1}.

In what follows, we will uncover all  gauge invariant models of the form 
(\ref{projectedaction}) which describe the free $(3/2,2)$ multiplet. We will set $\a_1=-\frac 1{16}$ to ensure canonical normalization.
\section{Minimal Models with Two Projectors}
~~~~There exist  three minimal theories which involve two superprojectors
and realize the free $(3/2,2)$ multiplet with $12+12$ off-shel degrees of 
freedom.

Consider first an action of the form (\ref{projectedaction}) which contains the 
projectors $\Pi^T_{3/2}$ and $\Pi^L_0$ only. All the coefficients turn out to be 
uniquely fixed: $\a_2=0$, $\a_3 =1/4$ and $\a_4 = 1/48$. As follows from (\ref{chil}), the action possesses the gauge invariance (\ref{gauge}) with the gauge parameter constrained by
\be
\Bar D^2D^\a L_\a =0~.
\label{o-m-con}
\ee
The constraint can be avoided by introducing a chiral scalar compensator $\s$, ${\Bar D}_\ad \s =0$, with the following gauge transformation
 \be
\d\s=-\frac 1{12}\Bar D^2 D^\a L_\a~.
\ee
With the compensator present, the action turns into the linearized action of old minimal  supergravity \cite{Gates1,Buch1}:
\bea
\label{oldmin}
{\cal S}_{\rm Old-Min} = \int d^8z \Big\{ H^{\un a}\Box(-\frac 13\P_{0}^L
+ \frac 12\P_{3/2}^T)H_{\un a} -i(\s-\bar\s)\pa_{\un a}H^{\un a}-3\s\bar\s \Big\}~.
\eea

Consider next an action  of the form (\ref{projectedaction}) which contains the projectors $\Pi^T_{3/2}$ and $\Pi^T_{1/2}$ only.  It is singled out by the conditions $\a_2 =0$, $\a_3 =1/4$ and $\a_4 = 1/16$. As follows from (\ref{lint}), the action possesses the gauge invariance (\ref{gauge}) provided the gauge parameter is constrained by
\be
D^\a\Bar D^2L_\a+\Bar D_{\dot\a}D^2\Bar L^{\dot\a} =0~.
\label{n-m-con}
\ee
The constraint can be eliminated at the cost of introducing a linear real compensator ${\cal U} = \Bar {\cal U}$,  ${\Bar D}^2 {\cal U} = 0$, with the gauge transformation
\be
\d \cu=\frac 14(D^\a\Bar D^2 L_\a+\Bar D_{\dot\a}D^2 \Bar L^{\dot\a})~.
\ee
As a result,  we end up with the linearized action of new minimal 
supergravity (see, e.g. \cite{Buch1})
\bea
\label{newmin}
{\cal S}_{\rm New-Min}= \int d^8z \Big\{H^{\un a}\Box(-\P_{1/2}^T +\frac 12
\P_{3/2}^T)H_{\un a} + \frac 12 {\mathcal U}  [D_\a,\Bar D_{\dot\a}]H^{\un a}
+\frac 32{\mathcal U}^2 \Big\}~.
\eea

We now turn to the case when only the projectors $\Pi^T_{3/2}$ and $\Pi^L_{
1/2}$ appear in the action (\ref{projectedaction}). This corresponds to the
choice  $\a_2 =0$, $\a_3 =1/12$ and $\a_4 = 1/48$. In accordance with 
(\ref{linl}), the action possesses the gauge invariance (\ref{gauge})
provided the gauge parameter is  constrained by
\be
D^\a\Bar D^2L_\a - \Bar D_{\dot\a}D^2\Bar L^{\dot\a} =0~.
\label{n-m2-con}
\ee
The constraint can be eliminated at the cost of introducing a linear real 
compensator  $ U = \Bar  U$,  ${\Bar D}^2  U = 0$, with the gauge transformation
\be
\d U=\frac{i}{12} (D^\a\Bar D^2 L_\a- \Bar D_{\dot\a}D^2 \Bar L^{\dot\a})~.
\ee
As a result,  we arrive at the following model
\bea
\label{2min}
{\cal S} =\int d^8z \Big\{H^{\un a}\Box (\frac 13\P_{ 1/2}^L +\frac 12\P_{3/2}^T)
H_{\un a} +U\pa_{\un a}H^{\un a} +\frac 32 U^2 \Big \}~,
\eea
which was derived in \cite{Buch2} and which partially prompted the present 
investigation.  The above  three models are known to be classically equivalent.  The  auxiliary action connecting the old minimal and new minimal models
is given in  \cite{Buch1}, while the auxiliary action relating the old minimal
model to (\ref{2min}) is given in \cite{Buch2}.

It remains to consider the case when the action (\ref{projectedaction})
involves the projectors $\Pi^T_{3/2}$ and $\Pi^T_{1}$ only.  It corresponds to 
the choice $\a_2 =1$,  $\a_3 = 3/4$ and $\a_4 = 1/16 $.  As is seen from 
(\ref{chit})  the gauge freedom is now constrained by
\be
\Bar D^2D_{(\a}L_{\b)} =0~.
\label{pathalogy}
\ee
Such gauge symmetry turns out to be too restrictive to correspond to a pure 
$(3/2, 2) $ multiplet. The simplest way to understand this is to look at the 
component structure of all the models so far discussed.  

With no compensators present,  we have to deal with a restricted set of  gauge transformations (\ref{gauge}) such that  $L_\a$ obeys some constraint which can be symbolically represented as follows (the precise position of spinor derivatives as well  as the index structure are not essential for our consideration at this point)
\be
\a \, {\Bar D}^2 \,D \,L  + \b \, D^2 \,{\Bar D}\, {\Bar L} =0~,
\ee
with $\a$ and $\b$ constant parameters.  This constraint does not impose any restrictions on the component fields of $L$ of order $\q^n$, where $n=0,1,2$. 
Therefore, we can gauge away  all the component fields of $H$ of order $\q^n$, with $n=0,1$, resulting with the Wess-Zumino gauge
\be
H_{\a \ad} ( \q, {\bar \q}) = \q^2 \, S_{\a \ad}  + {\bar \q}^2 \, {\bar S}_{\a \ad}
+  \q^\b {\bar \q}^\bd E_{\a \ad, \b \bd}  +\q^2 {\bar \q}^2 \, A_{\a \ad} + 
\mbox{fermions}~.
\label{WZ1}
\ee
Were $L_\a$  unconstrained, one could have gauged away $S_{\a\ad}$ completely.  When $L_\a$ is constrained as above,some part of $S_{\a\ad}$ can still be gauged away while the rest survives as an auxiliary field.

Let us concentrate on the component field $E_{\a \ad, \b \bd}$ which is to be identified with a spin-2 field. Its Lorentz irreducible  components are
\bea
h_{(\a \b) (\ad \bd)}~, \quad  h~ & \qquad &
\mbox{massless spin-2 field} ~; \cr
\O_{(\a \b)}~, \quad {\Bar \O}_{(\ad \bd)} & \qquad &
\mbox{gauge degrees of freedom}~.
\eea
The gauge superfield parameter $L_\a$ should not be over-constrained in the sense that it should contain enough component parameters to be able to gauge $\O_{(\a\b)}$ away. The relevant part of $L_\a$ occurs at the level ${\bar \q}^2 \q$ which is
\be
L_\a (\q, \bar \q )  ~\propto ~ {\bar \q}^2 \q_\a (f +i  \, g) + {\bar \q}^2
 \q^\b \U_{(\a \b)}~.
\ee
It is the parameter $\U_{(\a\b)}$ which allows us to gauge $\O_{(\a\b)}$ away.
In the case of  constraints (\ref{o-m-con}),  (\ref{n-m-con}) and (\ref{n-m2-con}), the gauge parameter $\U_{(\a\b)}$ remains completely arbitrary. On the other  hand, the constraint (\ref{pathalogy}) implies $\U_{(\a\b)}=0$, and hence there is no gauge freedom at all to eliminate $\O_{(\a\b)}$. As a result, the fourth model considered above does not describe the free $(3/2,2)$ multiplet. The same is actually true for any  action involving  the projector  $\Pi^T_{1}$.  This is why we will set $\a_2=0$ in what follows.

To get an idea of the main difference between the models (\ref{newmin}) and  
(\ref{2min}), which otherwise look very similar, it is worth pursuing the component analysis a bit further.  With the compensators present, we can choose a stronger Wess-Zumino  gauge than the one given in (\ref{WZ1}):
\be
H_{\a \ad} (\q, {\bar \q}) =  \q^\b {\bar \q}^\bd \Big( h_{ (\a \b) (\ad \bd)  }
+ \ve_{\a \b} \,\ve_{\ad \bd} \, h \Big) +\q^2 {\bar \q}^2 \, A_{\a \ad} ~+~ 
\mbox{fermions}~.
\label{grav}
\ee
The remaining independent gauge parameters are
\be
L_\a (\q, {\bar \q}) =  i \,{\bar \q}^\ad \, \z_{\a \ad} +{\bar 
\q}^2 \, \q_\a \,\Big( f +i\, g \Big) ~+~\mbox{fermions}~.
\label{L3}
\ee
Here the parameter $\z_{\a \ad} $ generates linearized spin-2 gauge 
transformations. The parameters $f$ and $g$ turn out to play rather different 
roles in the models (\ref{newmin}) and  (\ref{2min}).

In the new minimal model, the parameter $f$ can be used to gauge away 
the leading ($\q$-independent) component of the compensator $\cu$. After 
that,  $\cu$ possesses a single bosonic component  $\cg_{\a \ad}$,
\be
\cu  (\q, {\bar \q}) = \q^\a {\bar \q}^\ad \, \cg_{\a \ad}
~+~  \mbox{fermions}~,
\ee
which is the field strength of a gauge two-form, $\pa^a \cg_a =0$. The parameter $g$ generates local $\g_5$-transformations for which $A_m$ is a gauge field,
\be
\d A_m   \propto \pa_m\, g  ~.
\ee
The fields $G_m$ and $A_m$ appear in the action as follows
\be
\cs_{\rm New-Min} \propto
\int d^4 x \Big( A_m G^m + G_m G^m  \Big) ~,
\ee
with the local $\g_5$-invariance being manifest. It is only with the choice
$\a_4=1/16$, which is characteristic of the new minimal supergravity, that
no quadratic term $A^2$ is present in  the action.

In the model of \cite{Buch2}, the parameter $g$ can be used to gauge away the leading ($\q$-independent) component of the compensator $U$,
\be
U  (\q, {\bar \q}) = \q^\a {\bar \q}^\ad \, G_{\a \ad}
~+~\mbox{fermions}~,
\ee
with $G_{\a \ad}$ the field strength of a gauge two-form, $\pa^a G_a =0$. As to the parameter $f$, now it can be used to gauge away the trace component $h$ of the gravitational field, see (\ref{grav}),
\be
\d \, h = f~.
\ee
Setting $h=0$, the only remaining gauge symmetry is linearized general coordinate transformations generated by $\z_m$. The corresponding transformation law of $G_m$ is
\be
\d\, G_m ~\propto~ \pa^n (\pa_m \z_n - \pa_n \z_m )~.
\ee
In this formulation, the gravitational field is described by a {\it symmetric traceless} tensor
\be
h_{mn}~, \qquad h_{mn}= h_{nm} ~,
\qquad h^m{}_m =0
\ee
together with an {\it antisymmetric} tensor
\be
B_{mn}~, \qquad B_{mn} = -B_{nm}~,
\ee
which generates the field strength $G^m = {1 \over 2} \ve^{mnrs}\, \pa_n \, B_{rs}$.  The two fields, $h_{mn}$ and $B_{mn}$, can be combined  into a general traceless tensor  $h_{mn} + B_{mn}$. Essentially, what we end up with is a dual formulation for the linearized massless spin-2 action, in which 
the trace component $h$ of the gravitational field is traded for a gauge two-form.
\section{No Irreducible Models with Three Projectors}
~~~~We now turn to the study of actions with the projector $\Pi^T_{3/2}$
and two others. Since we have set $\a_2 =0$, there are three possible cases, using different combinations of the three projectors $\Pi^L_0$, $\Pi^L_{1/2}$ 
and $\Pi^T_{1/2}$. Technically, such models appear  to look like a sum of two 
of the minimal theories discussed in the previous section. Thus, we will simply present the gauge invariant actions with compensators included.

With  $\a_3\not=\frac 1{12},\frac 14$ and $\a_4=\frac 1{48}$ we arrive at
the  theory containing $\s$ and $U$:
\bea
\label{hot1}
{\cal S}_{\s,U}&=&\int d^8z\Big\{H^{\un a}\Box[-2(\a_3-\frac
1{12})\P_{0}^L -2(\a_3-\frac 14)\P_{1/2}^L+\frac 12\P_{3/2}^T]H_{\un
a}
\nonumber \\
&-&6[i(\a_3-\frac 1{12})(\s-\bar\s)+(\a_3-\frac 14)U]\pa_{\un a}H^{\un a}
\nonumber \\
&-&18(\a_3-\frac 1{12})\s\bar\s-9(\a_3-\frac 14)U^2 \Big\} ~.
\eea
If we set $\a_3=\frac 14$ and $\a_4\not=\frac 1{48}, \frac 1{16}$ we
are lead to the action:
\bea
\label{hot2}
{\cal S}_{\s,\,\cu}&=&\int d^8z\Big\{H^{\un a}\Box[+8(\a_4-\frac 1{16})
\P_{0}^L -24(\a_4-\frac 1{48})\P_{1/2}^T +\frac 12\P_{3/2}^T ]H_{\un a}
\nonumber \\
&-&12
[(\a_4-\frac 1{16})(\s+\bar\s) -(\a_4-\frac 1{48}) \cu]   [D_\a , \Bar D_{\dot\a}]
H^{\un a}
\nonumber \\
&+&72(\a_4-\frac 1{16})\s\bar\s +36(\a_4-\frac 1{48})\cu^2
\Big\}~.~~~~~~~~~~~
\eea
Finally, by setting $\a_3\not=\frac 14, \frac 1{12}$ and $\a_4=\frac 14\a_3$ 
we can construct an action containing only the linear compensators:
\bea
\label{hot3}
{\cal S}_{U,\,\cu}&=&
\int d^8z\Big\{ H^{\un a}\Box [-2(\a_3-\frac 14)\P_{1/2}^L
-6(\a_3-\frac 1{12})\P_{1/2}^T +\frac 12\P_{3/2}^T]H_{\un a}
\nonumber \\
&+&
3(\a_3-\frac 1{12}) {\cal U} [D_\a,\Bar D_{\dot\a}]H^{\un a}
-6(\a_3-\frac 14) U \pa_{\un a}H^{\un a}
\nonumber \\
&+&
9(\a_3-\frac 1{12}) {\cal U}^2 - 9(\a_3-\frac 14) U^2 \Big\}
~.
\eea
These actions describe $16+16$ off-shell degrees of freedom.

The above models prove to be equivalent as they are related to each other by superfield duality transformations. The following two dual actions connect (\ref{hot1}) 
to (\ref{hot3}) and (\ref{hot2}) to (\ref{hot3}), respectively:
\bea
\label{hot4}
{\cal S}_{\rm Aux}^{(1)}&=&\int d^8z \Big\{ H^{\un a}\Box [-2(\a_3-\frac 14)
\P_{1,\frac 12}^L -6(\a_3-\frac 1{12})\P_{1,\frac 12}^T +\frac 12
\P_{1,\frac 32}^T]H_{\un a}
\nonumber \\
&-&6(\a_3-\frac 14)U\pa_{\un a}H^{\un a}-9(\a_3-\frac 14)U^2
\\
&+& W[ 3(\a_3-\frac 1{12})[D_\a ,\Bar D_{\dot\a}]H^{\un a}
+\ 9 (\a_3-\frac 1{12})W - 18 (\a_3-\frac 1{12})(\s+\bar\s) ]\Big\}~;
\nonumber
\\
{\cal S}_{\rm {aux}}^{(2)}&=&\int d^8z \Big\{H^{\un a}\Box [-2(\a_3-\frac 14)
\P_{1,\frac 12}^L -6(\a_3-\frac 1{12})\P_{1,\frac 12}^T +\frac 12\P_{1,\frac 
32}^T]H_{\un a}
\nonumber \\
\label{hot5}
&+& 3 (\a_3-\frac 1{12}){\cal U}[D_\a , \Bar D_{\dot\a}]H^{\un a}
+9(\a_3-\frac 1{12}){\cal U}^2
 \\
&+& X[-6(\a_3-\frac 1{4})\pa_{\un a}H^{\un a}
- 9(\a_3-\frac 1{4})X -6i (\a_4-\frac 1{16})(\s-\bar\s)] \Big\} ~.
\nonumber
\eea
If we vary these actions with respect to the real unconstrained superfields $W$  and $X$,  we are led to (\ref{hot1}) and (\ref{hot2}).  If instead we vary the actions with respect to  $\s$, then $W$  and $X$ become real linear superfields,  and both (\ref{hot4}) and (\ref{hot5}) turn into  (\ref{hot3}).

Since the models (\ref{hot1}),  (\ref{hot2}) and (\ref{hot3}) are dually equivalent, it is sufficient to analyze the multiplet structure, say, of the model   (\ref{hot1}).  The important point here is that the leading ($\q$-independent) components of $\s$, $\Bar \s$ and $U$ are (pseudo) scalars of mass dimension $+1$.  If these fields cannot be algebraically gauged away, the theory contains propagating  
scalars and, therefore, does not describe a pure $(3/2,2)$ multiplet. In the 
Wess-Zumino gauge (\ref{grav}), there are only two  local parameters,  $f$ and $g$ in (\ref{L3}),  which can be used for gauging away the (pseudo) scalars under consideration.  These gauge parameters are not enough to do the job.

There might be a way out provided all the  dependence of  (\ref{hot1}) on the 
superfields $\s$, $\Bar \s$ and $U$ could be expressed via a single {\it  unconstrained} real superfield of the form $V= i (\a_3 -\frac{1}{12}) (\s- \Bar \s ) +  (\a_3-\frac{1}{4}) U$.  If this were possible, there would be enough gauge freedom to remove the dangerous leading component of $V$. A direct analysis shows this is not the case. Actually, in the action  (\ref{hot1}) one can trade the compensators  $\s$, $\Bar \s$ and $U$ for, say, a complex superfield $M=  i (\a_3 -\frac{1}{12}) \s + {1 \over 2}  (\a_3-\frac{1}{4}) U$ and its conjugate $\Bar M $ under the constraints
\bea
\Bar D^2 M=0~~,~~\Bar D^2 D_\a(M-\Bar M) =0~.
\eea
Now, one is  in a position to  gauge away the leading components of $M$ and $\Bar M$.  Unfortunately, $M$ still contains a transverse vector field, at the $\q\bar\q$ level, which is  the field strength of a two form gauge potential.  The gauge two-form enters the action as a propagating field.

Our conclusion is that the models (\ref{hot1}),  (\ref{hot2}) and (\ref{hot3}) do not describe an irreducible $(3/2, 2)$ multiplet.  In fact, at the full non-linear level, we have the following equivalences:
\begin{center}
(\ref{hot1}) = old minimal SUGRA coupled to a tensor multiplet;\\
(\ref{hot2}) = new minimal SUGRA coupled to a chiral multiplet;\\
(\ref{hot3}) = new minimal SUGRA coupled to a tensor multiplet.\\
\end{center}
This is in agreement with the the old result \cite{Siegel} that $16+16$ supergravity \cite{GGMW,LLO}  is reducible, and represents 
only a particular coupling of supergravity to a tensor multiplet.
\section{Non-minimal Models with Four Projectors}
~~~~It remains to analyze gauge invariant actions involving four projectors. With the standard choice $\a_1 = -\frac{1}{16}$, $\a_2=0$ and with the compensators included, the corresponding  action reads
\bea
 \cs
&=&\int d^8z\Big\{ H^{\un a}\Box [ (-2\a_3+8\a_4)\P_{0}^L +(\frac 12-2\a_3
)\P_{1/2}^L +(\frac 12-24\a_4)\P_{1/2}^T +\frac 12\P_{3/2}^T] H_{\un a} 
 \nonumber \\
&+&3 \Big[ i(-2\a_3+8\a_4)(\s-\bar\s) +(\frac 12-2\a_3)U\Big]  \,\pa_{\un a}
H^{\un a} -\frac 12(\frac 12-24\a_4)  \cu\, [D_\a , \Bar D_{\dot\a}]H^{\un a}
\nonumber \\
& +&9(-2\a_3+8\a_4)\s\bar\s -\frac 32(\frac 12-24\a_4)\cu^2
+\frac 92(\frac 12-2\a_3)U^2  \Big\}~,
\label{non-minimal}
\eea
and presents itself a sum of the three minimal theories derived  in section 3.  At first sight, such models look even more hopeless, in the sense of  describing  a pure $(3/2,2)$ multiplet,  than the ones considered in the previous section. 
Fortunately, the situation is not really that bad and there is a  way out. The point is that the compensators $\s$, $U$ and $\cal U$ can now be  arranged into a complex linear superfield $\S = a \, \s + b\, \cu  + i \,c \, U$, for some real coefficients $a,b$ and $c$, under the constraint
\be
{\Bar D}^2 \S =0~.
\ee
In the Wess-Zumino gauge (\ref{grav}), there is still enough gauge freedom to remove the dangerous $\q$-independent components of $\Sigma$ and $\Bar 
\S$ (as well as a  spinor component of $\S$ at the next-to-leading order); the 
remaining component fields of $\S$ and $\Bar \S$ enter the action simply as 
auxiliaries.  By expressing (\ref{non-minimal}) via $\S$ and $\Bar \S$ (along with implementing a field redefinition of the form $\S \to \S + \k {\Bar D}_\ad D_\a H^{\un a}$, for some parameter $\k$), one should end up with the linearized action of non-minimal supergravity parameterized by a complex parameter $n$ \cite{Gates2}.  Since the case of real-$n$  non-minimal supergravity is more familiar \cite{Gates1,Buch1}, we are going to  describe in more detail the procedure of how it occurs in the present framework.  The latter case turns out to correspond to a specific choice
\be
\S = a \, \s + b\ ( \cu  + 3 i  U)~,
\label{a-b}
\ee
in conjunction  with a special relationship between $\a_3$ and $\a_4$.

To express the action (\ref{non-minimal}) via $\S$ and $\Bar \S$, we should first match the $H^{\un a}$ couplings.  These terms necessarily take the form:
\bea
i (\S - \Bar \S)\, \pa_{\un a} H^{\un a} & =&
\Big\{ ia\, (\s-\bar\s) - 6b\,U \Big\} \, \pa_{\un a} H^{\un a}
 \\
\int d^8z (\S + \Bar \S ) [D_\a,\Bar D_{\dot\a}] H^{\un a}
&=& - 2 \int d^8z \Big\{i a \,  (\s-\bar\s) \pa_{\un a}H^{\un a}
-b \, {\cal U}  [D_\a,\Bar D_{\dot\a}]H^{\un a} \Big\}~. \nonumber
\eea
${}$Furthermore, since there is no coupling between $U$ and ${\cal U}$ the
kinetic terms for $\S$ must take the following form:
\bea
\int d^8z \S  \Bar\S &=& \int d^8z \Big\{a^2\s\bar\s+b^2( {\cal U}^2+9U^2)
 \Big\}  ~,
\nonumber \\
\int d^8z \Big\{\S^2 +\Bar\S^2 \Big \} & =&   2 b^2 \int d^8z \Big\{
{\cal U}^2 - 9U^2 \} ~.
\eea
Choosing the $\S$-dependent part of the  Lagrangian to be of the form
\bea
i (\S - \Bar\S) \pa_{\un a}H^{\un a} +\a (\S+\Bar\S ) [D_\a , \Bar D_{\dot\a}]
H^{\un a} +\b \S  \Bar\S  +\g(\S^2  + \Bar\S^2)~,
\label{sigma-dep}
\eea
we can readily get all the coefficients right. ${}$For the parameters $a$ and 
$b$ in (\ref{a-b}) we obtain
\bea
a=-\frac 14{{(-2\a_3+8\a_4)(\frac 12-2\a_3)}\over (-2\a_3+24\a_4)}~,
\qquad
b =  - \frac 12(\frac 12-2\a_3)~.
\eea
${}$Next, for the parameters $\a$ and $\g$ in (\ref{sigma-dep})
we get
\bea
\a = \frac 12{(\frac 12-24\a_4)\over (\frac 12-2\a_3)}~,
\qquad
\g={(2-2\a_3-72\a_4)\over(\frac 12-2\a_3)}~,
\eea
while for $\b$ there occur two different expressions,
\bea
\b={(-2\a_3+24\a_4)^2\over(-2\a_3+8\a_4)(\frac 12-2\a_3)^2}
={(72\a_4-2\a_3-1)\over (\frac 12-2\a_3)^2}~,
\eea
and these imply
\bea
\a_4 = {2\a_3\over 64\a_3+8}~.
\eea
Setting $\a_3=-{n+1 \over 8n}$  gives $\a_4={n+1 \over32}$.  This is the exact solution for linearized non-minimal supergravity given in \cite{Buch1}.
\vspace{5mm}

\noindent
{\bf Acknowledgements}  \\
SMK is grateful to Jim Gates, Warren Siegel
and Arkady Tseytlin  for warm hospitality at the University of Maryland,
C.N. Yang Institute for Theoretical Physics and Ohio State University,
where bits of this project were completed.  JP is grateful to the organizers of TASI 2003 for the hospitality at the University of Colorado where this manuscript was completed.  We would also like to thank I. L. Buchbinder and W. D. Linch, III for useful discussions.  

\end{document}